\documentclass[pdflatex,sn-basic, iicol]{sn-jnl}

\usepackage{graphicx}%
\usepackage{multirow}%
\usepackage{amsmath,amssymb,amsfonts}%
\usepackage{amsthm}%
\usepackage{mathrsfs}%
\usepackage{xcolor}%
\usepackage{textcomp}%
\usepackage{manyfoot}%
\usepackage{booktabs}%
\usepackage{caption}
\usepackage{subcaption}
\usepackage{algorithm}%
\usepackage{algorithmicx}%
\usepackage{algpseudocode}%
\usepackage{listings}%
\usepackage{braket}
\usepackage{float} 
\usepackage{threeparttable}
\begin{document}

\title[Quantum Autoencoders for Anomaly Detection in Cybersecurity]{Quantum Autoencoders for Anomaly Detection in Cybersecurity}


\author[1]{\fnm{Rohan}\sur{Senthil}}\email{rohansenthil.biz@gmail.com}

\author*[1]{\fnm{Swee Liang} \sur{Wong}}\email{wong\_swee\_liang@htx.gov.sg}

\affil[1]{\orgdiv{Disruptive Technologies Office}, \orgname{Home Team Science \& Technology Agency}, \orgaddress{\street{1 Stars Avenue}, \city{Singapore}, \postcode{138507}, \state{Singapore}, \country{Singapore}}}


\abstract{Anomaly detection in cybersecurity is a challenging task, where normal events far outnumber anomalous ones with new anomalies occurring frequently. Classical autoencoders have been used for anomaly detection, but struggles in data-limited settings which quantum counterparts can potentially overcome. In this work, we apply Quantum Autoencoders (QAEs) for anomaly detection in cybersecurity, specifically on the BPF-extended tracking honeypot (BETH) dataset. QAEs are evaluated across multiple encoding techniques, ansatz types, repetitions, and feature selection strategies. Our results demonstrate that an 8-feature QAE using Dense-Angle encoding with a RealAmplitude ansatz can outperform Classical Autoencoders (CAEs), even when trained on substantially fewer samples. The effects of quantum encoding and feature selection for developing quantum models are demonstrated and discussed. In a data-limited setting, the best performing QAE model has a F1 score of 0.87, better than that of CAE (0.77). These findings suggest that QAEs may offer practical advantages for anomaly detection in data-limited scenarios.}




\maketitle
\section{Introduction}\label{Introduction}

In today’s fast-evolving digital environment, cyber threats continue to increase in both sophistication and frequency. Traditional security measures often struggle to keep up with the relentless influx of attacks \citep{Hussainetal2020}. Such cyber threats often manifest as anomalous activities in the system. Thus, means of finding these threats can be done through anomaly detection. Anomaly detection is a key component of data analysis that focuses on identifying data points or events that deviate significantly from the norm. It plays a vital role in areas such as cybersecurity and fraud detection, where spotting unusual patterns can be critical for uncovering potential threats and preventing breaches \citep{Tscharkeetal2023}.  

However, the current cybersecurity landscape is marked by an unprecedented growth in the volume, velocity, and complexity of security-related data \citep{Alluhaibi2024}. This results in normal cyber-activities far outnumbering anomalous ones, making detecting of anomalies a challenging task. This is evidenced in cybersecurity-related datasets which are highly imbalanced, where normal activities far outnumber anomalous ones. \citep{Highnametal2021}. Furthermore, the rapidly evolving landscape for cybersecurity and the continuously varying activity data \citep{Petersenetal2015} means that a significant challenge for supervised anomaly detection techniques is maintaining large, high-quality, labeled, and relevant datasets \citep{Mirskyetal2018}. Moreover, as threats evolves, so will the nature of how normal and anomalous data are represented.

Previous works suggest the use of machine learning methods such as Support Vector Machines and deep learning techniques such as Convolutional Neural Networks to distinguish regular from suspicious activities.\citep{Sabaetal2021}. Other methods include autoencoders which are trained solely on 'normal' data\citep{Mirskyetal2018}\citep{Neloyetal2024}\citep{Patraetal2022} due to the aforementioned scenario of limited anomalous data. Classical Autoencoders (CAEs) learn the encoding distribution of 'normal' data, so that when an anomalous data is reconstructed using a CAE trained only on 'normal' data, its resultant reconstruction error will be much higher as the anomaly does not conform to the distribution. \citep{Wuetal2024}. 

The quantum autoencoder (QAE), introduced by \cite{Romeroetal2017}, extends the CAE concept to quantum systems. QAEs can compress and decompress quantum states, flagging outliers based on either reconstruction errors or by measuring fidelity from a SWAP test \citep{Liuetal2018}\citep{Herretal2021}\citep{Tscharkeetal2023}\citep{Frehneretal2025}. QAEs have potential to provide advantageous data efficiency and expressibility due to the ability of QAEs to operate in higher dimensions compared to classical counterparts.

Our research proposes the use of Quantum Autoencoders for anomaly detection in cybersecurity, leveraging the intrinsic expressibility and efficiency of quantum circuits. In this work, we demonstrated that QAEs, trained solely on normal data from the BETH dataset, outperform their classical counterparts in data-limited settings in detecting anomalies. The choice of encoding and ansatz are also discussed.

\section{Related Work}\label{Related_Work}
\subsection{Classical Autoencoders}\label{Classical_Autoencoders}
CAEs have been used for anomaly detection in cybersecurity applications, particularly effective in scenarios where anomalous data is scarce and explicit labeling is impractical. These models function by learning compact latent representations of normal system behavior and can accurately reconstruct similar input data with low error under such conditions. In contrast, anomalous inputs that deviate significantly from learned patterns yield higher reconstruction errors, enabling their identification \citep{Afzaletal2025}.

Recent progress has extended beyond basic autoencoders to more advanced architectures. Variational Autoencoders introduce regularization by constraining latent representations to follow prior distributions, while Adversarial Autoencoders leverage generative adversarial networks to align encodings with arbitrary priors \citep{Mohamed2023}\citep{Larsenetal2016} Nonetheless, achieving the right balance between reconstruction quality and generalization remains a challenge: excessive compression can hinder accurate reconstruction of normal data, while overly complex models risk overfitting and reduced sensitivity to anomalies \citep{Mohamed2023}.

\subsection{Quantum Autoencoders}\label{Quantum_Autoencoders}
QAEs represent a powerful quantum machine learning paradigm that extends CAEs by leveraging uniquely quantum properties such as superposition and entanglement to enhance anomaly detection capabilities \citep{Romeroetal2017}. In high-energy physics, for instance, QAEs have been shown to outperform CAEs on data generated from the Large Hadron Collider, achieving superior reconstruction performance with substantially fewer training samples and parameters \citep{Ngairangbametal2022}. This drastic reduction in parameter requirements while maintaining strong performance underscores a key advantage of quantum methods, particularly in data-constrained settings.

Early work by Liu et al. \citep{Liuetal2018} provided evidence of the potential speedup and resource efficiency for quantum computing based anomaly detection methods. Previous work demonstrates that QAEs, when augmented with quantum gradient descent for training, are much less dependent on the number of training samples and reach optimal reconstruction performance with small training datasets \citep{Ngairangbametal2022}. This efficiency extends to model complexity, where QAEs with approximately 10 parameters can outperform classical autoencoders with around 1000 parameters in anomaly detection capabilities \citep{Ngairangbametal2022}.

Recent reports on applying QAEs to time-series anomaly detection in cybersecurity data demonstrate consistently better performance over CAEs, while using fewer parameters and training iterations \citep{Frehneretal2025}. In Frehner et al's work, it explored the effects of different ansatz and measurement techniques (SWAP-based vs Reconstructed Mean Squared Error). However, there is still a limited understanding of how different quantum encoding and feature selection strategies impact the performance of QAEs, indicating opportunities for further optimization and specialization \citep{Arazetal2024}. This suggests that the development of specialized QAE variants is still in its early stages, with substantial potential for domain-specific optimizations and architectural approaches.

\subsubsection{Quantum Encoding}\label{Quantum_Encoding}
Quantum encoding techniques are data preprocessing steps that transform classical data into quantum states that can be processed by quantum circuits. The method of encoding data is also known as applying a quantum feature map \citep{Schuldetal2019}.

The choice of encoding method significantly impacts the expressivity of quantum machine learning models, affecting their ability to capture complex patterns in data and their computational efficiency. Different encoding techniques create distinct quantum state spaces and correlation structures, which can either enhance or limit a model's capacity to learn from specific types of data. Each encoding method creates distinct topological structures that can preserve or distort the original data relationships, directly affecting downstream tasks such as clustering and classification \citep{Vlasicetal2023}. This indicates that the performance of quantum machine learning models is highly dependent on the choice of encoding, which we will explore in this work. 

Thus in this work, we applied QAE to detect anomalies, using the BETH dataset and the performance is compared against CAE. We explore the effects of sample sizes, feature selection and quantum encoding techniques on the model performance. In our experiments, we implement and evaluate the following encoding techniques: (1) Amplitude Encoding, (2) Angle Encoding, (3) Dense-Angle Encoding, (4) Efficient SU2 Encoding. Various ansatz and feature selection are investigated, and their effects on the QAE model's performance are demonstrated. 

\section{Dataset}\label{Dataset}
\subsection{The BETH Dataset}\label{BETH_Dataset}
The BPF-Extended Tracking Honeypot (BETH) dataset is a purpose-built cybersecurity dataset designed to support machine learning research in network security, which comprises tabular event-level network traffic data collected from real-world honeypots \citep{Highnametal2021}. It provides a large collection of network traffic logs that include both benign activity and malicious intrusions, making it a strong candidate for testing anomaly detection methods. Overall, the dataset consists of \textit{8,004,918 events} gathered from honeypots deployed on a major cloud service provider, spanning approximately \textit{five noncontiguous hours} of operation on a major cloud provider. For benchmarking and discussion, an initial subset of the process logs was divided into training, validation, and testing sets with a rough 60/20/20 split. The split considered host origin, log volume, and activity type, with attacks appearing only in the test set. The distribution of normal and anomalous events in each subset is shown in Table~\ref{tab:beth_distribution}.  

\begin{table*}[t]
\centering
\captionsetup{}
\caption{Distribution of normal and anomalous events in the BETH dataset.}
\label{tab:beth_distribution}
\begin{tabular}{lccc}
\toprule
\textbf{Dataset} & \textbf{Normal} & \textbf{Anomalous} & \textbf{Percentage Anomalous} \\
\midrule
Training   & 761,875 & 1,269   & 0.2\%  \\
Validation & 188,181 & 786     & 0.4\%  \\
Testing    & 17,508  & 171,459 & 90.7\% \\
\bottomrule
\end{tabular}
\end{table*}

The practical relevance of the BETH dataset extends beyond academic research, as it reflects real-world network conditions and attack patterns that cybersecurity professionals encounter in operational environments. This real-world applicability makes it particularly valuable for evaluating how QAEs perform on practical cybersecurity challenges.

\subsection{Data Preprocessing}\label{Data_Preprocessing}
Data preprocessing and feature engineering is performed on the BETH dataset. Feature transformations were applied in accordance with the recommendations by \citeauthor{Highnametal2021}:

\begin{itemize}
    \item \texttt{ProcessId} and \texttt{ParentProcessId}: mapped to 0 if in \{0, 1, 2\}, else 1.
    \item \texttt{UserId}: mapped to 0 if \textless 1000, else 1.
    \item \texttt{MountNameSpace}: mapped to 0 if equal to 4026531840, else 1.
    \item \texttt{ReturnValue}: mapped to 0 if zero, 1 if positive, and 2 if negative.
    \item \texttt{Timestamp} column dropped due to non-temporal modeling.
    \item \texttt{EventName} dropped due to one-to-one mapping with \texttt{EventId}.
    \item \texttt{ProcessName}, \texttt{ThreadId}, and \texttt{EventId} encoded using target encoding.
    \item \texttt{StackAddresses} and \texttt{HostName} were dropped.
    \item \texttt{Args}: each row contains a list of up to 5 dictionaries, each with keys \texttt{name}, \texttt{type}, and \texttt{value}. These were flattened into 15 new columns and encoded using target encoding.
\end{itemize}

All categorical features were processed via target encoding. Target encoding was chosen due to high-cardinality present in the data which causes other methods such as One Hot Encoding to result in overfitting. No normalization was applied for all features except for \texttt{ArgsNum}, which was scaled using standard scaling. In total, the transformed BETH dataset contained 24 features.

\section{Experiments}\label{Experiments}

QAE was implemented in \texttt{Qiskit} and simulated on local hardware. The primary computing environment consisted of an \texttt{11th Gen Intel\textsuperscript{\textregistered} Core\texttrademark{} i7-11800H} processor (16 threads), \texttt{Intel\textsuperscript{\textregistered} UHD Graphics (TGL GT1)}, and 16 GB of RAM, running \texttt{Ubuntu 24.10}. In addition, \texttt{Google Colab L4 GPUs} were also employed in parallel to train multiple model configurations. 

The performance of the trained QAE model is evaluated against CAE by comparing the precision, accuracy, recall, F1-score, AUROC as well as detection statistics. The model is solely trained on the train dataset split with no anomalous data present in it while evaluation is done against the test dataset provided by the BETH dataset.

\subsection{Quantum Autoencoder Architecture}\label{QAE_Architecture}

\subsubsection*{Circuit Structure}\label{Circuit_Structure}

The typical QAE consists of three conceptual layers: (1) \textit{Input}, (2) \textit{Bottleneck}, and (3) \textit{Output}.  

\begin{figure}[H]
    \centering
    \includegraphics[width=1.1\linewidth]{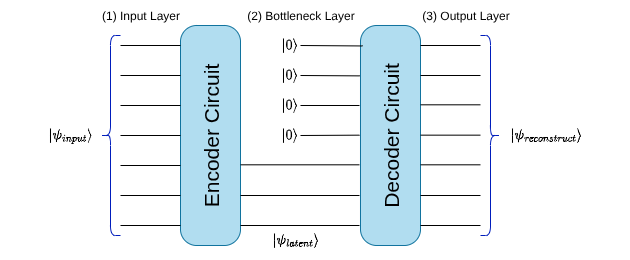}
    \captionsetup{justification=raggedright,singlelinecheck=false}
    \caption{Diagram of a typical QAE}
    \label{fig:qae_daigram}
\end{figure}

\begin{enumerate}
    \item \textit{Input Layer (Encoder):}  
    The input state $\ket{\psi}$, defined over $n$ qubits, is initialized and passed through a parameterized circuit. This circuit acts as the encoder, compressing the quantum state into a reduced subspace of $n-k$ latent qubits. The resulting state takes the form
    \[
        \ket{\psi}_{\text{latent}} \otimes \ket{0}^{\otimes k}
    \]
    where $\ket{\psi}_{\text{latent}}$ contains the $n-k$ latent qubits, while the remaining $k$ qubits are discarded.
    
    \item \textit{Bottleneck Layer:}  
    The $k$ discarded qubits are removed from the circuit, leaving only the compressed subspace or latent qubits as the representation.
    
    \item \textit{Output Layer (Decoder):}  
     To reconstruct the original state,  a second parameterized circuit, acting as the decoder, is applied jointly on the latent qubits and $k$ qubits initialized in $\ket{0}$ to approximate the original state. However, as mentioned later, our implementation of SWAP test means that we do not need to carry out reconstruction of the input and can hence disregard the second parameterized circuit.
     
\end{enumerate}

\subsubsection*{State Comparison via SWAP Test}\label{SWAP_Test}

To assess reconstruction fidelity without explicitly performing the full reconstruction, the $k$ discarded or trash qubits are each paired with $k$ reference qubits (predefined quantum states serving as fidelity benchmarks) and compared via a SWAP test \citep{Asaokaetal2025}. An additional auxillary qubit is also introduced which is used for measurement tasks. 

The SWAP test proceeds by applying CNOT gates between the auxiliary qubit with each pair of trash-reference qubits being compared (i.e. in Figure~\ref{fig:ansatzs} $q2$ compared to $q4$ and $q3 $ compared to $q5$). After executing the circuit $M$ times, the auxiliary qubit is measured. Let $L$ denote the number of times the auxiliary qubit is measured to be $\ket{1}$. The corresponding similarity score is defined as
\begin{equation}
    S = 1 - \frac{2}{M} L.
\end{equation}

Maximizing $S$ corresponds to maximizing the fidelity between the two states, which is equivalent to ensuring that the input and reconstructed states are as close as possible. Thus, the training objective is to minimize the cost function $\tfrac{2}{M}L$.  

Therefore, the anomaly score, $A$, is defined as
\begin{equation}
A=1 - S
\end{equation}
where $S$ is the fidelity score obtained from the SWAP test. Since $S \in [0,1]$, the anomaly score $A$ also lies in $[0,1]$, with $A=0$ indicating a perfectly normal instance and $A=1$ representing a maximally anomalous instance.

\subsubsection*{Ansatz and Training}\label{Ansatz_Training}

For efficiency, we restricted our experiments to 8 features, corresponding to at most 8 qubits. Different encoding techniques require varying numbers of qubits to represent the same number of features; accordingly, the distribution between latent and trash qubits is technique-dependent.

\begin{figure}[H]
    \centering
    \begin{subfigure}{0.40\textwidth}
        \centering
        \includegraphics[width=\textwidth]{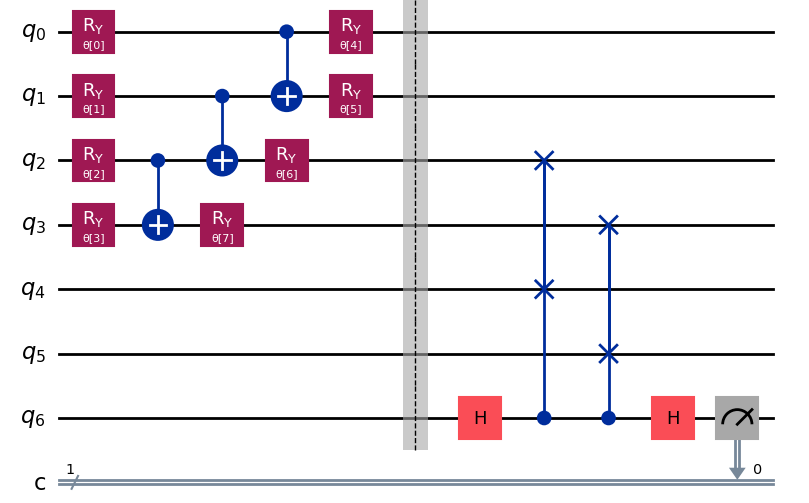}
        \caption{RealAmplitude Ansatz Circuit}
        \label{fig:realamp}
    \end{subfigure}
    \hfill
    \begin{subfigure}{0.40\textwidth}
        \centering
        \includegraphics[width=\textwidth]{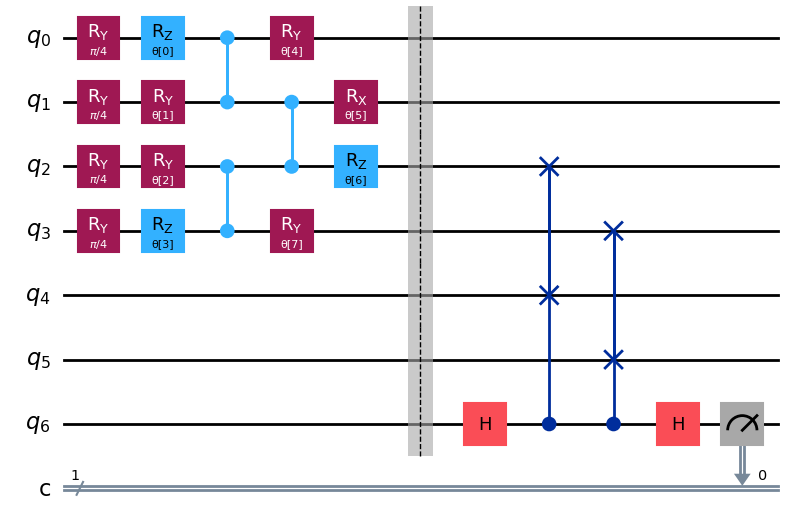}
        \caption{PauliTwoDesign Ansatz Circuit}
        \label{fig:paulitwo}
    \end{subfigure}
    \caption{Comparison of two ansatz circuits: (a) RealAmplitude and (b) PauliTwoDesign. The features are encoded into $q0$ to $q3$ where $q0$ and $q1$ are the latent qubits and $q2$ and $q3$ are the trash qubits, $q4$ to $q5$ represent the reference qubits and $q6$ represents the auxiliary qubit.}
    \label{fig:ansatzs}
\end{figure}

Multiple parameterized ansatz are explored, such as \texttt{RealAmplitudes} (RA) and \texttt{PauliTwoDesign} (Pauli), which are illustrated in Figures~\ref{fig:realamp} and \ref{fig:paulitwo}. The figures illustrate the different ansatz when Dense-Angle encoding is applied which encodes 8 features into 4 qubits.

For optimization, we employed the COBYLA algorithm, a gradient-free optimizer well-suited for quantum hardware, and performed 60 training iterations on different sample sizes with a batch size of 64.

\subsection{Encoding Techniques}\label{Encoding_Techniques}

\subsubsection{Amplitude Encoding}\label{Amplitude_Encoding}
Amplitude encoding encodes data into the amplitudes of a quantum state. It represents a normalized classical $N$-dimensional data vector, as the amplitudes of a $n$-qubit quantum state.

Consider a dataset $X=x^{1} + ... + x^{i} ... + x^{m}$ containing m samples, where each sample is a vector with n features, then the amplitude encoding can be represented as follows:

\begin{equation}
\ket{\psi^{i}} = \sum_{j=1}^{n}x_{j}^{i} \ket{j}
\end{equation}

\subsubsection{Angle Encoding}\label{Angle_Encoding}
For angle encoding, each feature of the data sample is encoded into the rotation angle of each qubit where the data is scaled as $\phi \in [0,\pi]$. For the angle encoding of $x^i$, we require $n$ quantum bits to complete the encoding \citep{Shietal2024}. The encoding is represented as:
\begin{equation}
\ket{\psi^{i}} = \overset{n}{\underset{j=1}{\otimes}}R(x_j^{i})
\end{equation}
where, $R$ is a single qubit $R_y$ rotation gate. 

\subsubsection{Dense-Angle Encoding}\label{Dense_Angle_Encoding}
Dense-Angle encoding is a combination of angle encoding and phase encoding. Phase encoding is similar to the angle encoding described above. The phase angle of a qubit is a real-valued angle $\phi$ about the $z$-axis from the $+x$-axis. Data is mapped with a phase rotation. 

Dense-Angle allows two feature values to be encoded in a single qubit: one with a $y$-axis rotation angle ($R_y$ gate), and the other with a $z$-axis rotation angle ($R_z$ gate). Again, data is scaled according to $\phi \in [0,\pi]$. Encoding two data features into one qubit results in a $2$ times reduction in the number of qubits required.

\subsubsection{Efficient SU2 Encoding}\label{Efficient_SU2_Encoding}
An encoding scheme from \texttt{Qiskit}, Efficient SU2, which involves entanglement between the qubits is investigated as well. In our implementation, data is scaled where $\phi \in [0,\pi]$. It is made of $y$-axis rotations, $z$-axis rotations along with entanglements of neighboring qubits. This circuit can, for example, encode 8 features on only 2 qubits.

With the Efficient SU2 encoding we evaluated 16- and 24-feature inputs instead of 8, as 8 features would only allow 1 trash and 1 latent qubit that can result in loss of information due to overcompressing of data.

\subsection{Classical Autoencoder Architecture}\label{Classical_Autoencoder_Architecture}

For benchmarking, we also implemented a CAE. (see Supplementary Information B for more details).

The model was compiled using the \textit{Adam optimizer} and the \textit{Mean Absolute Error (MAE)} as the loss function. The anomaly score is also defined as the \textit{MAE} between the reconstructed data and the original data.

\begin{figure}[H]
    \centering
    \includegraphics[width=0.40\textwidth]{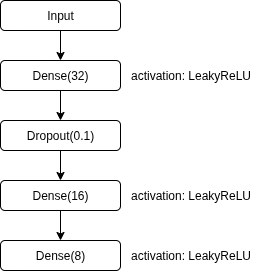}
    \captionsetup{}
    \caption{Encoder Architecture for 24 input feature model}
    \label{fig:classical_autoencoder}
\end{figure}

The encoder architecture for a 24 input feature model, as shown in Figure~\ref{fig:classical_autoencoder}, was determined by experimenting with different hyperparameters. The decoder architecture is the reverse of the encoder. Anomaly scores are derived from the reconstruction error between input and output data.

\subsection{Thresholding Strategy}\label{Thresholding_Strategy}
Anomalies are determined with a threshold defined as:
\begin{equation}
\text{Threshold} = \mu_{\text{error}} + 2 \cdot \sigma_{\text{error}}
\end{equation}
where $\mu_{\text{error}}$ and $\sigma_{\text{error}}$ represent the mean and standard deviation of the reconstruction errors on the
training (normal) data, respectively. When the anomaly score exceeds this defined threshold, the data is considered anomalous. Once again, we experimented with different thresholding techniques
and determined the above to be the most optimal.

\subsection{Feature Selection Strategies}\label{Feature_Selection_Strategies}

The BETH dataset has 24 features, which requires a maximum of 24 qubits when angle encoding is employed. Due to computational resource constraints, only 8 features are selected in this work. These features are selected through different means and dimensionality reduction techniques. This includes using the latent space produced by the classical autoencoder as well as other methods such as Principal Component Analysis, Mutual Information, ANOVA, Kendall's and arbitrary selection methods such as First-N, and Least Correlated features.

\section{Results}\label{Results}

For the QAE, we explored different encoding techniques, feature selection strategies, ansatz types, and ansatz repetitions. We also experimented with increasing feature counts where feasible. Additionally, various entanglement techniques (linear, reverse-linear, full) for the ansatz were tested. Our choice of ansatz, encoding strategies and feature selection strategies are evaluated on a training size of 1000 samples.

Our key findings are quantitatively expressed using the F1 score as $\text{mean} \pm \sigma$, where $\sigma$ represents the standard deviation of observed F1 scores. The results are summarized as follows:

\begin{itemize}
    \item \textit{Feature selection:} Using Amplitude and Dense-Angle Encoding as the basis for evaluating the feature selections strategies, F1 scores are compared: First-N (0.798 $\pm$ 0.063) comparatively performed the best consistently. Comparatively, other strategies such as ANOVA (0.44 $\pm$ 0.350), Kendall's (0.460 $\pm$ 0.262) and Least Correlated (0.380 $\pm$ 0.300) varied significantly across different encoding techniques.\\

    \begin{minipage}{\linewidth}
    \begin{center}
    \small
    \captionsetup{}
    \captionof{table}{Mean $\pm$ $\sigma$ of F1 scores for different feature selection strategies using Amplitude and Dense-Angle encoding.}
    \begin{tabular}{l c}
    \hline
    \textbf{Feature Selection Strategy} & \textbf{F1 Score} \\
    \hline
    First-N & 0.798 $\pm$ 0.063 \\
    Autoencoder-based reduction & 0.695 $\pm$ 0.069 \\
    PCA (dimensionality reduction) &  0.628 $\pm$ 0.143 \\
    PCA (feature filtering) &  0.685 $\pm$ 0.045 \\
    Mutual Information & 0.705 $\pm$ 0.065 \\
    ANOVA &  0.440 $\pm$ 0.350 \\
    Kendall’s & 0.460 $\pm$ 0.262 \\
    Least Correlated & 0.380 $\pm$ 0.300 \\
    \hline
    \end{tabular}
    \end{center}
    \label{tab:feature_selection_f1}
    \end{minipage}
    
    \item \textit{Encoding Techniques:} Table~\ref{tab:dimension_reduced_performance} and Table~\ref{tab:firstn_performance} describes the model performance for different encoding techniques. Using autoencoder feature reduction and First-N feature selection as the basis to compare the F1 scores between models of different encoding techniques, we observe that Amplitude (0.705 $\pm$ 0.060) and Dense-Angle (0.771 $\pm$ 0.069) achieve the strongest overall performance. In comparison, Angle (0.709 $\pm$ 0.083) has greater variance in model performance and EfficientSU2 (0.675 $\pm$ 0.032) does not perform as well.

    \begin{minipage}{\linewidth}
    \begin{center}
    \small
    \captionsetup{}
    \captionof{table}{Comparison of QAE models across different encoding techniques where CAE was used for dimensionality reduction. Row in bold indicate top performing model.}
    \begin{tabular}{|c|c|c|c|c|}
    \hline
    \textbf{Encoding} & \textbf{Ansatz} & \textbf{Reps} & \textbf{F1} & \textbf{AUROC} \\
    \hline
    Amplitude & RA & 1 & 0.76 & 0.914 \\
    Amplitude & RA & 3 & 0.58 & 0.413 \\
    Amplitude & Pauli & 1 & 0.71 & 0.859 \\
    Amplitude & Pauli & 3 & 0.76 & 0.110 \\
    Angle & RA & 1 & 0.69 & 0.537 \\
    Angle & RA & 3 & 0.69 & 0.583 \\
    Angle & Pauli & 1 & 0.53 & 0.552 \\
    Angle & Pauli & 3 & 0.71 & 0.120 \\
    Dense & RA & 1 & 0.73 & 0.895 \\
    Dense & RA & 3 & 0.71 & 0.085 \\
    \textbf{Dense} & \textbf{Pauli} & \textbf{1} & \textbf{0.86} & \textbf{0.918} \\
    Dense & Pauli & 3 & 0.71 & 0.131 \\
    24F EffSU2 & RA & 1 & 0.63 & 0.814 \\ 
    24F EffSU2 & RA & 3 & 0.70 & 0.103 \\ 
    24F EffSU2 & Pauli & 1 & 0.66 & 0.453 \\ 
    24F EffSU2 & Pauli & 3 & 0.71 & 0.683 \\ 
    \hline
    \end{tabular}
    \label{tab:dimension_reduced_performance}
    \end{center}
    \end{minipage}
    
    \begin{minipage}{\linewidth}
    \begin{center}
    \small
    \captionsetup{}
    \captionof{table}{Performance metrics for First-N features across different encoding and ansatz configurations. Row in bold indicates the top-performing model.}
    \begin{tabular}{|c|c|c|c|c|}
    \hline
    \textbf{Encoding} & \textbf{Ansatz} & \textbf{Reps} & \textbf{F1} & \textbf{AUROC} \\
    \hline
    Amplitude & RA & 1 & 0.72 & 0.872 \\ 
    Amplitude & RA & 3 & 0.75 & 0.899 \\ 
    Amplitude & Pauli & 1 & 0.72 & 0.790 \\ 
    Amplitude & Pauli & 3 & 0.64 & 0.424 \\ 
    Angle & RA & 1 & 0.74 & 0.864 \\ 
    Angle & RA & 3 & 0.85 & 0.896 \\ 
    Angle & Pauli & 1 & 0.73 & 0.900 \\ 
    Angle & Pauli & 3 & 0.73 & 0.835 \\ 
    \textbf{Dense} & \textbf{RA} & \textbf{1} & \textbf{0.86} & \textbf{0.960} \\ 
    Dense & RA & 3 & 0.86 & 0.955 \\ 
    Dense & Pauli & 1 & 0.72 & 0.918 \\ 
    Dense & Pauli & 3 & 0.72 & 0.096 \\ 
    24F EffSU2 & RA & 1 & 0.63 & 0.814 \\ 
    24F EffSU2 & RA & 3 & 0.70 & 0.103 \\ 
    24F EffSU2 & Pauli & 1 & 0.66 & 0.453 \\ 
    24F EffSU2 & Pauli & 3 & 0.71 & 0.683 \\ 
    \hline
    \end{tabular}
    \label{tab:firstn_performance}
    \end{center}
    \end{minipage}

\end{itemize}

Based on these results, the best performing model comprises the following: \textit{(a) first 8 features of data as input, (b) Dense Angle encoding and (c) RealAmplitude ansatz}. This model is evaluated against a similarly trained CAE.

\begin{figure}[H]
    \centering
    \scriptsize
    \begin{subfigure}[b]{\linewidth}
        \centering
        \includegraphics[width=0.6\linewidth]{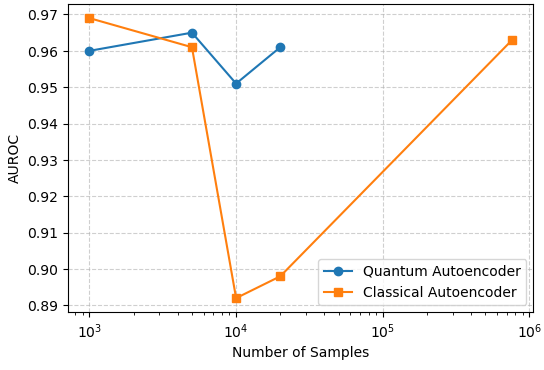}
        \caption{AUROC vs. Samples}
        \label{fig:auroc_samples}
    \end{subfigure}
    
    \vspace{0.5em}
    
    \begin{subfigure}[b]{\linewidth}
        \centering
        \includegraphics[width=0.6\linewidth]{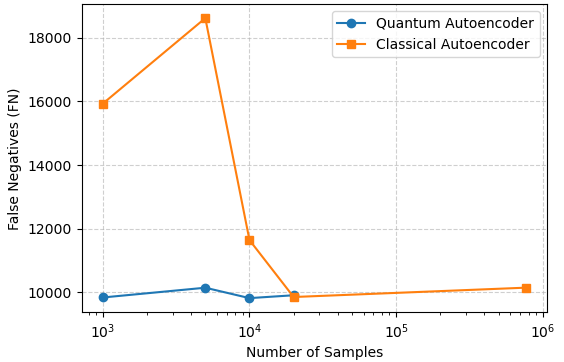}
        \caption{FNs vs. Samples}
        \label{fig:fn_samples}
    \end{subfigure}
    
    \vspace{0.5em}
    
    \begin{subfigure}[b]{\linewidth}
        \centering
        \includegraphics[width=0.6\linewidth]{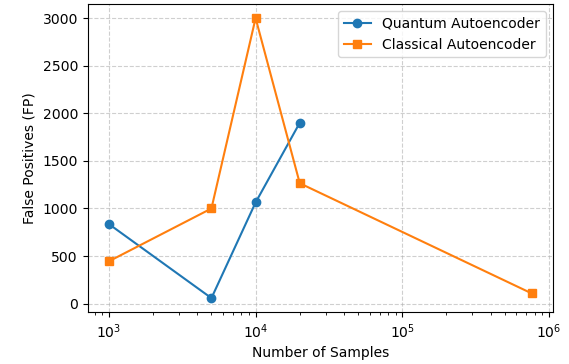}
        \caption{FPs vs. Samples}
        \label{fig:fp_samples}
    \end{subfigure}
    
    \captionsetup{}
    \caption{Performance comparison of QAEs and CAEs trained with 1,000, 5,000, 10,000, and 20,000 samples. CAE also includes the full training dataset. QAE was not trained with the full dataset due to resource constraints.}
    \label{fig:performance_comparison}
\end{figure}

We illustrate in Figure~\ref{fig:performance_comparison} the effects of sample size on the performance of both the CAE and QAE. To accommodate for variations between subsets of data, random data subsets consisting 5,000 samples are evaluated. The sample size is chosen based on preliminary tests which indicates the highest model performance to sample size ratio for the QAE. The F1 scores (0.82 $\pm$ 0.040) demonstrate slight variations as different samples are used for training but still performs the best compared to the other models. Greater details of the results for the different sample sizes and repetitions for the QAE can be found in Supplementary Information F.

Overall, the trained QAE model outperforms the classical counterpart when trained on the same number of samples. In Table~\ref{tab:qae_cae} different configurations of the CAE is compared against our best performing QAE model.

\begin{table*}
\centering
\small
\captionsetup{}
\caption{Comparison of CAE and QAE. All averages show mean ± $\sigma$ across repeated 5k random subsets.}
\begin{tabular}{|l|c|c|c|c|c|}
\hline
\textbf{Model} & \textbf{Features} & \textbf{Samples} & \textbf{Subset} & \textbf{F1} & \textbf{AUROC} \\
\hline
QAE & 8  & 5,000   & First Run & 0.87 & 0.965 \\
QAE & 8  & 5,000   & Average   & 0.82 ± 0.040 & 0.909 ± 0.033 \\
CAE & 8  & 5,000   & Average   & 0.77 ± 0.017 & 0.901 ± 0.044 \\
CAE & 8  & 761,875 & Full   & 0.66 & 0.964 \\
CAE & 24 & 5,000   & Average   & 0.77 ± 0.025 & 0.862 ± 0.066 \\
CAE & 24 & 761,875 & Full      & 0.87 & 0.963 \\
\hline
\end{tabular}
\label{tab:qae_cae}
\end{table*}

\begin{figure*}[hbtp]
    \centering
    \begin{subfigure}[b]{0.32\textwidth}
        \centering
        \includegraphics[width=\textwidth]{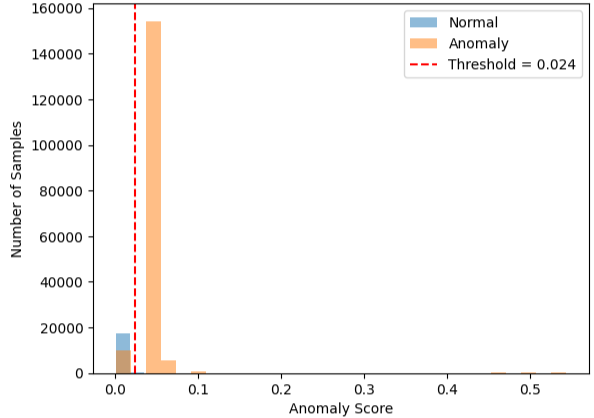}
        \captionsetup{font=footnotesize}
        \caption{Score Distribution of Best CAE simulated locally}
        \label{fig:CAE_Distribution}
    \end{subfigure}
    \hfill
    \begin{subfigure}[b]{0.32\textwidth}
        \centering
        \includegraphics[width=\textwidth]{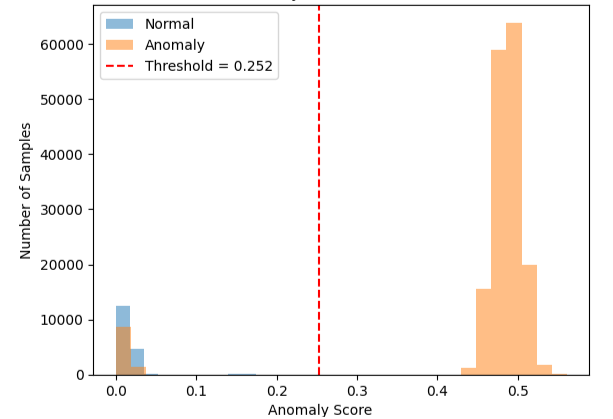}
        \captionsetup{font=footnotesize}
        \caption{Score Distribution of Best QAE on a noiseless simulation}
        \label{fig:QAE_Distribution}
    \end{subfigure}
    \hfill
    \begin{subfigure}[b]{0.32\textwidth}
        \centering
        \includegraphics[width=\textwidth]{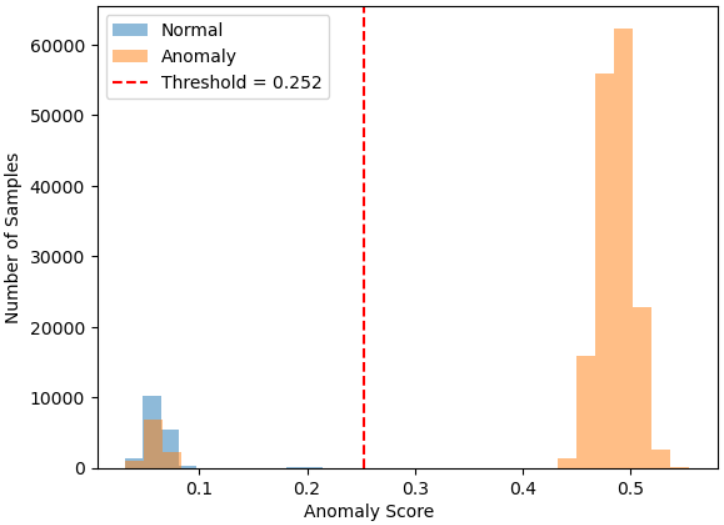}
        \captionsetup{font=footnotesize}
        \caption{Score Distribution of Best QAE with IBM Brisbane noise profile}
        \label{fig:noise_distribution}
    \end{subfigure}

    \captionsetup{}
    \caption{Score distributions of (a) the CAE simulated locally, (b) the QAE on a local noiseless simulation, and (c) the QAE on a local simulation with IBM Brisbane noise profile. While the CAE shows minimal separation, the QAE demonstrates a clear distinction between normal and abnormal data. The noise slightly shifts normal points, with no significant shift for anomalous data.}
    \label{fig:all_distributions}
\end{figure*}

To compare the anomaly score distributions as seen in Figure~\ref{fig:CAE_Distribution} and Figure~\ref{fig:QAE_Distribution}, we define \textit{separation} as 
\begin{equation}
\text{separation} = \frac{\text{mode}_{\text{anomaly}} - \text{mode}_{\text{normal}}}{\text{mode}_{\text{anomaly}}}
\end{equation}
The CAE trained on all 24 features achieves a separation score of \textit{0.84}, whereas the best-performing QAE attains a higher score of \textit{0.97}. This substantial improvement underscores the QAE’s effectiveness in anomaly detection and its superior ability to distinguish normal from anomalous data

Simulated noise was also added to our experiments to simulate operation on real-world quantum hardware, specifically using the noise profile of the IBM Brisbane quantum computer. On the first batch of testing data, the anomaly scores increased by roughly 2.83x due to noise induced variation in measurement outcomes. Despite this increase, the model's overall performance remains unaffected. Only minor shifts in the anomaly score distribution are observed (see Figure~\ref{fig:noise_distribution}. Due to the strong separation between the anomaly scores of normal data and that of anomalous data, these minor shifts do not affect the results, and all other evaluation metrics remain unchanged.

\section{Discussion}\label{Discussion}

In our experiments, when appropriate features and encoding techniques were selected, QAEs consistently outperformed a CAE trained on the same sample size.
Specifically, comparison between the QAE and the 8-feature input CAE shows that, at 5,000 training samples, the QAE achieves higher F1 score (+5\%) compared to the CAE trained on the same subset. Notably, the QAE trained on the first 5,000 samples outperforms the CAE trained on the full dataset, highlighting the quantum model’s efficiency in data-limited scenarios.

The anomaly score distributions and separation (0.97) further demonstrate the QAE’s strength in identifying anomalous data. One possible explanation for the QAE’s strong performance with limited data is that in higher-dimensional Hilbert space, subtle correlations and feature interactions become more apparent. Superposition and entanglement, which potentially enable richer encodings, also allow the model to simultaneously capture complex patterns and suppress irrelevant variations, reducing the need for large datasets compared to CAEs.

\begin{figure*}
    \centering
    \begin{subfigure}[t]{0.23\textwidth}
        \centering
        \includegraphics[width=\linewidth]{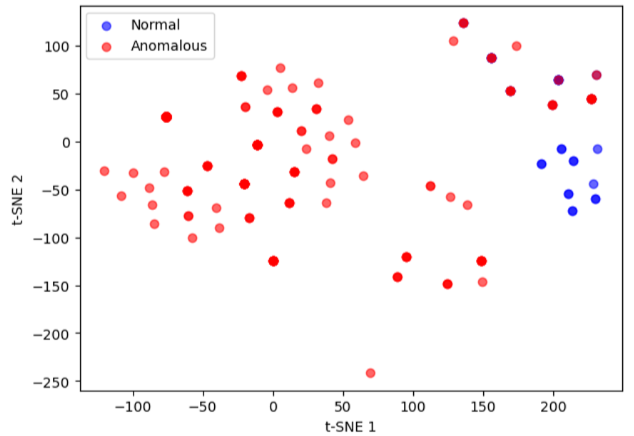}
        \captionsetup{font=footnotesize}
        \caption{First-N tSNE}
    \end{subfigure}
    \hfill
    \begin{subfigure}[t]{0.23\textwidth}
        \centering
        \includegraphics[width=\linewidth]{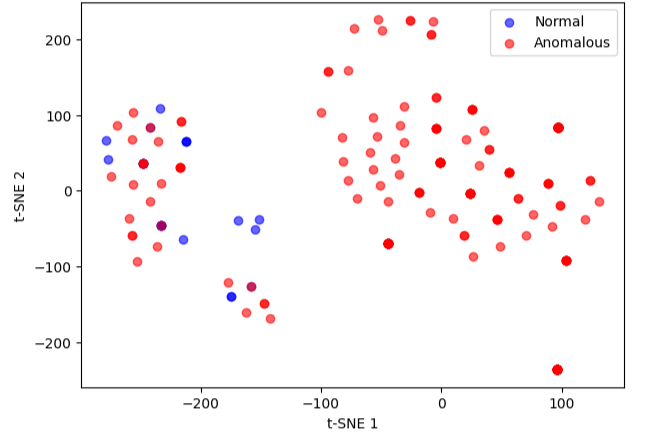}
        \captionsetup{font=footnotesize}
        \caption{PCA-filter tSNE}
    \end{subfigure}
    \hfill
    \begin{subfigure}[t]{0.23\textwidth}
        \centering
        \includegraphics[width=\linewidth]{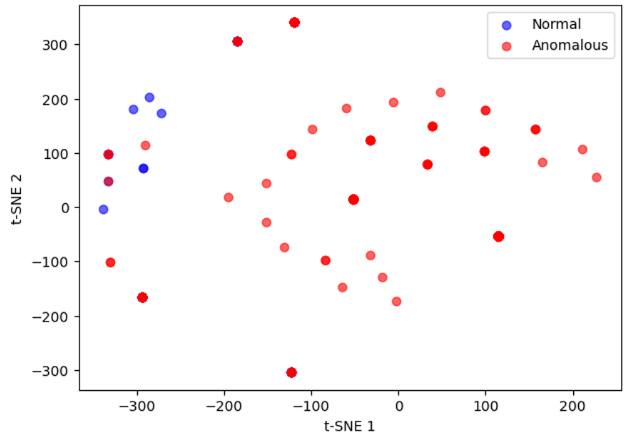}
        \captionsetup{font=footnotesize}
        \caption{Mutual Information tSNE}
    \end{subfigure}
    \hfill
    \begin{subfigure}[t]{0.23\textwidth}
        \centering
        \includegraphics[width=\linewidth]{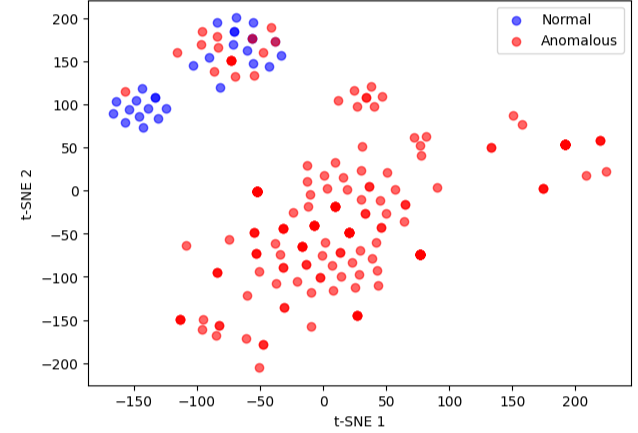}
        \captionsetup{font=footnotesize}
        \caption{ANOVA tSNE}
    \end{subfigure}
    
    \vspace{0.5em} 

    \begin{subfigure}[t]{0.23\textwidth}
        \centering
        \includegraphics[width=\linewidth]{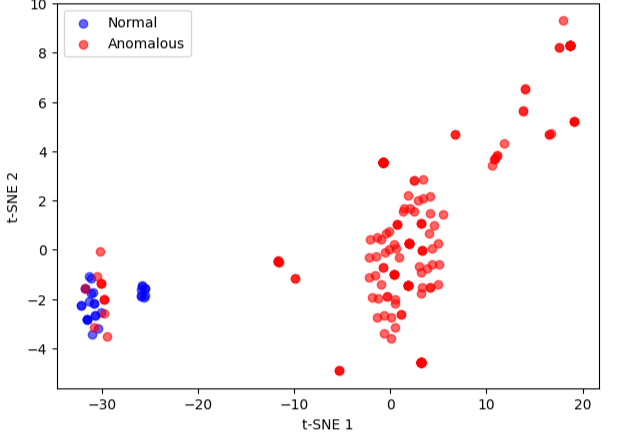}
        \captionsetup{font=footnotesize}
        \caption{Kendall's tSNE}
    \end{subfigure}
    \hfill
    \begin{subfigure}[t]{0.23\textwidth}
        \centering
        \includegraphics[width=\linewidth]{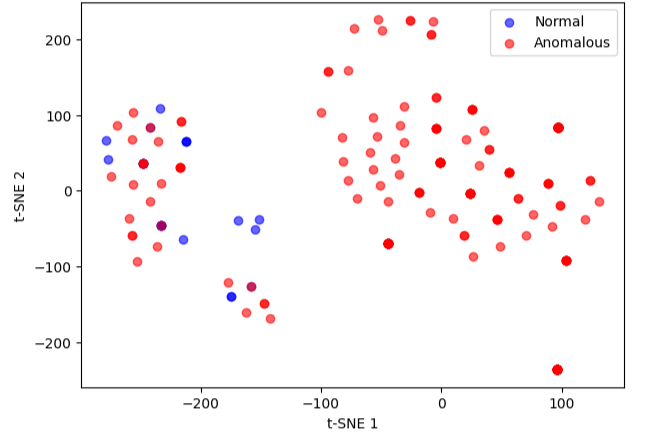}
        \captionsetup{font=footnotesize}
        \caption{Least Correlated tSNE}
    \end{subfigure}
    \hfill
    \begin{subfigure}[t]{0.23\textwidth}
        \centering
        \includegraphics[width=\linewidth]{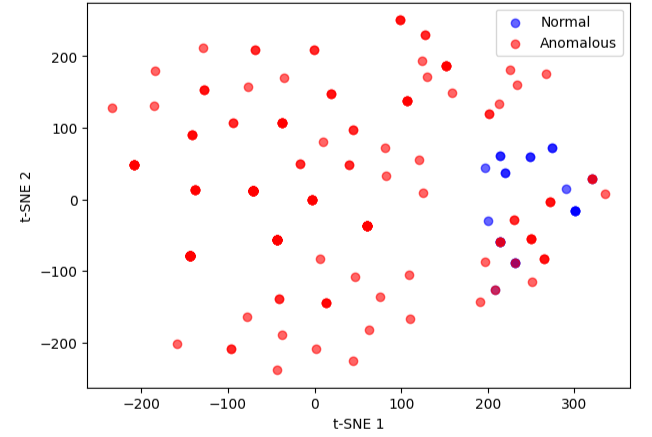}
        \captionsetup{font=footnotesize}
        \caption{HighestVar tSNE}
        \end{subfigure}
    \hfill
    \hfill
    \begin{subfigure}[t]{0.23\textwidth}
        \centering
        \includegraphics[width=\linewidth]{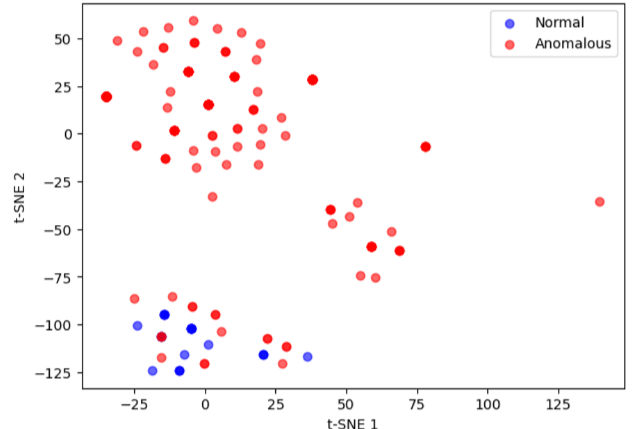}
        \captionsetup{font=footnotesize}
        \caption{BalancedVar tSNE}
    \end{subfigure}

    \captionsetup{}
    \caption{tSNE plots of various feature selection strategies.}
    \label{fig:8_diagrams}
\end{figure*}

For feature selection, the First-N approach achieved the best results. As seen in Figure~\ref{fig:8_diagrams}, t-SNE visualizations for First-N revealed clear and well-separated clusters. We hypothesize that this stems from its higher and more diverse feature variance ($\mu$: 0.44, $\sigma$: 0.52), which facilitates separation in anomaly scores.

In comparison, PCA ($\mu$: 0.22, $\sigma$: 0.44) and Mutual Information (MI) ($\mu$: 0.13, $\sigma$: 0.16) exhibited lower variance, resulting in weaker t-SNE clustering and less discriminative features. ($\mu$: 0.22, $\sigma$: 0.41) and Kendall correlation-based selection ($\mu$: 0.22, $\sigma$: 0.41) included some very low-variance features alongside higher-variance ones. Least Correlated ($\mu$: 0.27, $\sigma$: 0.44) captured diverse but uneven variance, yielding intermediate performance.

At the same time, a diverse but high-variance selection ($\mu$: 0.66, $\sigma$: 0.49) leads to poor model performance.

To explore the effect of feature variance on model performance, we implemented a variance-balanced feature selection strategy. Specifically, we first computed the variance of each feature across the dataset, then sorted features from lowest to highest variance. From this sorted list, we selected a predefined number of features from three variance categories: low, mid-range, and high variance. For 24 features, each group has 8 features. The lowest group was drawn from the bottom of the distribution (lowest 8), the middle group from the central third of the distribution (middle 8), and the high group from the top (first 8), ensuring that both low- and high-variance features were explicitly represented. For instance, in our 8-feature subset, 2 features were chosen from the lowest-variance group, 4 from the mid-range, and 2 from the highest-variance group. The variance-balanced selection ($\mu$: 0.33, $\sigma$: 0.56) improved clustering relative to PCA, MI, and correlation-based methods by explicitly including features with both low and high variance. Though the model train with (a) first 5000 training samples, (b) Dense Angle encoding and (c) RealAmplitude ansatz still did not surpass First-N with a F1 score of 0.81 and a separation score of 0.89. These findings suggest that feature selection strategies that maintain variance diversity (mean and range) are most effective for anomaly detection in this setting.

As shown in our results, encoding techniques affect the performance of the QAEs. Both Amplitude and Dense-Angle encoding \textit{consistently perform better} (Amplitude F1: 0.705 $\pm$ 0.060, Dense-Angle F1: 0.771 $\pm$ 0.069) than Angle encoding (F1:0.709 $\pm$ 0.083) which is less consistent and Efficient SU2 encoding (F1: 0.675 $\pm$ 0.032) which has lower performance. Comparing performance based on QAE trained with data using only first-N features, Amplitude and Angle encodings show comparable overall performance, but they excel in different aspects: Amplitude encoding more effectively reduces FPs, while Angle encoding is better at minimizing FNs. In contrast, Dense-Angle strikes a balance between these metrics and generally outperforms both, likely due to its incorporation of both rotations and phase information. Our results also indicate that Efficient SU2 is more sensitive to the choice of ansatz and the number of repetitions and Angle is more sensitive to the choice of features. For Angle encoding specifically, this sensitivity may arise from the lack of inherent entanglement in the encoding scheme, whereas Amplitude and Dense-Angle includes entanglement between qubits, enhancing their ability to capture multi-feature correlations and hence experiences less variation with the choice of ansatz and features, performing well even with limited data.

\section{Future Work}\label{Future_Work}

Our results indicate that QAEs surpass CAEs in anomaly detection in instances with limited training data. However, this advantage is highly dependent on appropriate feature selection and encoding strategies, with model performance largely impacted by them. Future research should focus on developing systematic frameworks and algorithms for identifying optimal features and selecting suitable encoding techniques. Considering constraints in qubit numbers in current hardware, further work is needed to design feature selection methods that are generalisable across different datasets. Similarly, there is still room for investigation into the impact of different encoding techniques across datasets, data types, and circuit architectures. Finally, extending this study to a broader range of datasets would provide stronger insights into how data characteristics influence both CAE and QAE performance.

Moreover, while QAEs show clear potential, resource constraints of current quantum hardware still remains. Further development in quantum hardware and algorithms would be required to achieve performance gains that outweigh the additional cost and complexity in implementing these QAEs.

\section{Conclusion}
This study explored the potential of QAEs for anomaly detection, benchmarking their performance against CAEs on the BETH dataset. Our findings highlight that, when combined with effective feature selection and carefully chosen encoding techniques, QAEs can outperform their classical counterparts. We also note the ability of QAEs to separate anomalies from normal samples as evident by a separation of \textit{0.97}. Experiments also demonstrate the sensitivity of QAEs to both input representation and circuit design—underscoring the need for principled approaches to encoding and feature engineering in quantum machine learning.

Overall, our work contributes an early empirical comparison between CAEs and QAEs in a practical anomaly detection task, while opening avenues for systematic methods in feature and encoding selection. With further refinements, QAEs may well be a robust alternative to classical methods.

\bmhead{Supplementary information}

BETH dataset details, performance statistics of different models and their respective architectures can be found in the Supplementary information.

\bmhead{Acknowledgements}
The authors acknowledge the Home Team Science and Technology Agency for financial support.

\bmhead{Author Contributions}
R.S developed the main conceptual ideas, set up and conducted the 
experiments and performed the data analysis. S.L.W supervised the project. Both authors wrote and reviewed the manuscript.

\section*{Declarations}

\bmhead{Funding}
Not Applicable

\bmhead{Data Availability}
Code used in this work is available on reasonable request. The data is publicly available as referenced in the text.

\bmhead{Competing Interests}
The authors declare no competing interests.

\bibliography{references}

\begin{thebibliography}{23}
\providecommand{\natexlab}[1]{#1}
\providecommand{\url}[1]{{#1}}
\providecommand{\urlprefix}{URL }
\providecommand{\doi}[1]{\url{https://doi.org/#1}}
\providecommand{\eprint}[2][]{\url{#2}}
 \bibcommenthead

\bibitem[{Abdulla et~al.(2020)Abdulla, Mohamed, and Razali}]{Hussainetal2020}
Abdulla A, Mohamed A, Razali S (2020) A review on cybersecurity: Challenges and emerging threats. In: International Conference on Networking, Information Systems \& Security (NISS), pp 1--7, \doi{10.1145/3386723.3387847}, \urlprefix\url{https://doi.org/10.1145/3386723.3387847}

\bibitem[{Afzal et~al.(2025)Afzal, Gaggero, and Asplund}]{Afzaletal2025}
Afzal Z, Gaggero G, Asplund M (2025) Towards privacy-preserving anomaly-based intrusion detection in energy communities. arXiv preprint \doi{10.48550/arXiv.2502.19154}, \urlprefix\url{https://arxiv.org/abs/2502.19154}, {\href{https://arxiv.org/abs/2502.19154}{{arXiv:2502.19154}}} {[cs.CR]}

\bibitem[{Alluhaibi(2024)}]{Alluhaibi2024}
Alluhaibi R (2024) Quantum machine learning for advanced threat detection in cybersecurity. International Journal of Safety and Security Engineering 14(3):875--883. \doi{10.18280/ijsse.140319}, \urlprefix\url{https://doi.org/10.18280/ijsse.140319}

\bibitem[{Araz and Spannowsky(2024)}]{Arazetal2024}
Araz JY, Spannowsky M (2024) The role of data embedding in quantum autoencoders for improved anomaly detection. arXiv preprint \doi{10.48550/arXiv.2409.04519}, \urlprefix\url{https://arxiv.org/abs/2409.04519}, {\href{https://arxiv.org/abs/2409.04519}{{arXiv:2409.04519}}} {[quant-ph]}

\bibitem[{Asaoka and Kudo(2025)}]{Asaokaetal2025}
Asaoka H, Kudo K (2025) Quantum autoencoders for image classification. Quantum Machine Intelligence 7(2). \doi{10.1007/s42484-025-00297-x}, \urlprefix\url{http://dx.doi.org/10.1007/s42484-025-00297-x}

\bibitem[{Frehner and Stockinger(2025)}]{Frehneretal2025}
Frehner R, Stockinger K (2025) Applying quantum autoencoders for time series anomaly detection. Quantum Machine Intelligence \doi{10.1007/s42484-025-00285-1}, \urlprefix\url{https://doi.org/10.1007/s42484-025-00285-1}

\bibitem[{Herr et~al.(2021)Herr, Obert, and Rosenkranz}]{Herretal2021}
Herr D, Obert B, Rosenkranz M (2021) Anomaly detection with variational quantum generative adversarial networks. Quantum Science and Technology 6(4):045004. \doi{10.1088/2058-9565/ac0d4d}, \urlprefix\url{https://doi.org/10.1088/2058-9565/ac0d4d}

\bibitem[{Highnam et~al.(2021)Highnam, Arulkumaran, Hanif, and Jennings}]{Highnametal2021}
Highnam K, Arulkumaran K, Hanif Z, et~al (2021) Beth dataset: Real cybersecurity data for unsupervised anomaly detection research. Proceedings of the Conference on Applied Machine Learning for Information Security (CAMLIS) \urlprefix\url{https://ceur-ws.org/Vol-3095/paper1.pdf}

\bibitem[{Larsen et~al.(2016)Larsen, Sønderby, Larochelle, and Winther}]{Larsenetal2016}
Larsen ABL, Sønderby SK, Larochelle H, et~al (2016) Autoencoding beyond pixels using a learned similarity metric. arXiv preprint \doi{10.48550/arXiv.1512.09300}, \urlprefix\url{https://arxiv.org/abs/1512.09300}, {\href{https://arxiv.org/abs/1512.09300}{{arXiv:1512.09300}}} {[cs.LG]}

\bibitem[{Liu and Rebentrost(2018)}]{Liuetal2018}
Liu N, Rebentrost P (2018) Quantum machine learning for quantum anomaly detection. Physical Review A 97(4):042315. \doi{10.1103/PhysRevA.97.042315}, \urlprefix\url{https://doi.org/10.1103/PhysRevA.97.042315}

\bibitem[{Mirsky et~al.(2018)Mirsky, Doitshman, Elovici, and Shabtai}]{Mirskyetal2018}
Mirsky Y, Doitshman T, Elovici Y, et~al (2018) Kitsune: An ensemble of autoencoders for online network intrusion detection. arXiv preprint arXiv:180209089 \doi{10.48550/arXiv.1802.09089}, \urlprefix\url{https://arxiv.org/abs/1802.09089}

\bibitem[{Mohamed(2023)}]{Mohamed2023}
Mohamed M (2023) Comparative evaluation of vaes, vae-gans and aaes for anomaly detection in network intrusion data. EMITTER International Journal of Engineering Technology 11(2):160--173. \doi{10.24003/emitter.v11i2.817}, \urlprefix\url{https://emitter.pens.ac.id/index.php/emitter/article/view/817}

\bibitem[{Neloy and Turgeon(2024)}]{Neloyetal2024}
Neloy AA, Turgeon M (2024) A comprehensive study of auto-encoders for anomaly detection: Efficiency and trade-offs. Machine Learning with Applications 17:100572. \doi{10.1016/j.mlwa.2024.100572}, \urlprefix\url{https://www.sciencedirect.com/science/article/pii/S2666827024000483}

\bibitem[{Ngairangbam et~al.(2022)Ngairangbam, Spannowsky, and Takeuchi}]{Ngairangbametal2022}
Ngairangbam VS, Spannowsky M, Takeuchi M (2022) Anomaly detection in high-energy physics using a quantum autoencoder. Physical Review D 105(9):095004. \doi{10.1103/PhysRevD.105.095004}, \urlprefix\url{https://doi.org/10.1103/PhysRevD.105.095004}

\bibitem[{Patra et~al.(2022)Patra, Sethi, and Behera}]{Patraetal2022}
Patra K, Sethi R, Behera D (2022) Anomaly detection in rotating machinery using autoencoders based on bidirectional lstm and gru neural networks. Turkish Journal of Electrical Engineering and Computer Sciences 30(7):2649--2662. \doi{10.55730/1300-0632.3870}, \urlprefix\url{https://journals.tubitak.gov.tr/elektrik/issues/elk-22-30-7/elk-30-7-19-2111-40.pdf}

\bibitem[{Petersen(2015)}]{Petersenetal2015}
Petersen R (2015) Data mining for network intrusion detection: A comparison of data mining algorithms and an analysis of relevant features for detecting cyber-attacks. Semantic Scholar \urlprefix\url{https://api.semanticscholar.org/CorpusID:64401231}

\bibitem[{Romero et~al.(2017)Romero, Olson, and Aspuru-Guzik}]{Romeroetal2017}
Romero J, Olson JP, Aspuru-Guzik A (2017) Quantum autoencoders for efficient compression of quantum data. Quantum Science and Technology 2(4):045001. \doi{10.1088/2058-9565/aa8072}, \urlprefix\url{http://dx.doi.org/10.1088/2058-9565/aa8072}

\bibitem[{Saba et~al.(2021)Saba, Sadad, Rehman, Mehmood, and Javaid}]{Sabaetal2021}
Saba T, Sadad T, Rehman A, et~al (2021) Intrusion detection system through advance machine learning for the internet of things networks. IT Professional 23(2):58--64. \doi{10.1109/MITP.2020.3016287}, \urlprefix\url{https://ieeexplore.ieee.org/document/9335650}

\bibitem[{Schuld and Killoran(2019)}]{Schuldetal2019}
Schuld M, Killoran N (2019) Quantum machine learning in feature hilbert spaces. Physical Review Letters 122(4):040504. \doi{10.1103/PhysRevLett.122.040504}, \urlprefix\url{https://doi.org/10.1103/PhysRevLett.122.040504}

\bibitem[{Shi et~al.(2024)Shi, Situ, and Zhang}]{Shietal2024}
Shi M, Situ H, Zhang C (2024) Hybrid quantum neural network structures for image multi-classification. Physica Scripta 99(5):056012. \doi{10.1088/1402-4896/ad3e3d}, \urlprefix\url{https://dx.doi.org/10.1088/1402-4896/ad3e3d}

\bibitem[{Tscharke et~al.(2023)Tscharke, Issel, and Debus}]{Tscharkeetal2023}
Tscharke K, Issel S, Debus P (2023) Semisupervised anomaly detection using support vector regression with quantum kernel. In: IEEE International Conference on Quantum Computing and Engineering (QCE), pp 611--620, \doi{10.1109/QCE57702.2023.00075}, \urlprefix\url{https://ieeexplore.ieee.org/document/10313835}

\bibitem[{Vlasic and Pham(2023)}]{Vlasicetal2023}
Vlasic A, Pham A (2023) Understanding the mapping of encode data through an implementation of quantum topological analysis. Quantum Information and Computation 23(13--14):1091--1104. \doi{10.26421/QIC23.13-14-2}, \urlprefix\url{http://dx.doi.org/10.26421/QIC23.13-14-2}

\bibitem[{Wu et~al.(2024)Wu, Fu, Zhu, Zhang, Xie, and Li}]{Wuetal2024}
Wu JY, Fu H, Zhu M, et~al (2024) Quantum circuit autoencoder. Physical Review A 109(3):032623. \doi{10.1103/PhysRevA.109.032623}, \urlprefix\url{https://journals.aps.org/pra/abstract/10.1103/PhysRevA.109.032623}

\end{thebibliography}

\end{document}